\documentclass{aa}  

\usepackage{xcolor}
\usepackage[colorlinks=true,linkcolor=blue,citecolor=blue]{hyperref}%
\usepackage{graphicx}
\usepackage{comment}

\usepackage{multirow}

\usepackage{subcaption}
\usepackage{txfonts}

\usepackage{float}

\usepackage[para,online,flushleft]{threeparttable}

\begin{document}

   \title{Beyond single tracers: CNN-based inference of galaxy mass profiles from combined gas and stellar kinematics}

   \author{J. Expósito-Márquez
          \inst{1}\fnmsep\inst{2}, A. Di Cintio\inst{2}\fnmsep\inst{1}, C. Brook\inst{2}\fnmsep\inst{1}, J. Sarrato-Alós\inst{1}\fnmsep\inst{2} and A.V.Macci\`{o}\inst{3}\fnmsep\inst{4}\fnmsep\inst{5}}

   \institute{Instituto de Astrofísica de Canarias (IAC), Calle Via Láctea s/n, E-38205 La Laguna, Tenerife, Spain
         \and
             Universidad de La Laguna, Avda. Astrofísico Fco. Sánchez s/n, E-38206 La Laguna, Tenerife, Spain
        \and
            New York University Abu Dhabi, PO Box 129188 Abu Dhabi, United Arab Emirates
        \and
            Center for Astro, Particle and Planetary Physics, New York University Abu Dhabi
        \and
            Max Planck Institute f\"{u}r Astronomie, K\"{o}nigstuhl 17, D-69117 Heidelberg, Germany}

   \date{Received x xx, xxxx; accepted x xx, xxxx}

 
  \abstract
   {}
   {We investigate whether combining gas and stellar kinematic maps provides measurable advantages in recovering galaxy mass profiles, compared to using single-component maps alone. While traditional methods struggle to integrate multi-tracer data effectively, we test whether deep learning models can leverage this joint information.}
   {We develop a probabilistic convolutional neural network (CNN) framework trained and tested on mock galaxy kinematic maps from multiple cosmological simulation suites. Our model is trained on gas-only, stars-only, and combined gas+stellar velocity maps, allowing direct comparison of performance across tracers. To assess robustness, we include simulations with differing feedback models and galaxy properties.}
   {Combining gas and stellar maps reduces the dispersion in the inferred mass profiles by up to a factor of $\sim$1.5 compared to models using either tracer independently. The CNN architecture effectively captures complementary information from the two components. However, we find limitations in generalizing between simulation suites, with reduced performance when applying models trained on one suite to galaxies from another.}
   {}

   {}


   \titlerunning{Inferring DM distributions with gas and stars}
   \authorrunning{J. Expósito-Márquez, A. Di Cintio, C. Brook, J. Sarrato-Alós, Andrea Macci\`{o}}

   \maketitle
%

\section{Introduction}
\label{sec:introduction}

The $\Lambda$CDM model provides a self-consistent framework for predicting galaxy formation from the initial density fluctuations observed in the cosmic microwave background (CMB), primarily through the influence of dark matter (DM), whose nature remains unknown. It successfully accounts for key properties of galaxies, including their abundance, clustering, morphologies, and evolution \citep[e.g.][]{vogelsberger14,schaye15}. However, observations on sub-Mpc scales present challenges to the model, raising the question of whether these discrepancies arise from baryonic physics, non-standard DM properties, or require a revision of the standard cosmological paradigm.

Understanding how  mass is distributed within galaxies is crucial for testing the CDM paradigm. Analysis of the rotation velocity of gas in low surface brightness galaxies, for example, allows to derive and fit their underlying DM distribution \citep[e.g.][]{moore94,gentile04,deblok08,lelli16,katz17}. On the other side, in pressure-supported galaxies which are devoid of gas, such as the dwarf spheroidal galaxies (dSphs) found within the Local Group, the kinematic information on which dynamical modeling relies on, comes from the line-of-sight velocity distribution of their stellar component. A variety of methods have been employed on dwarf galaxies to derive their central DM density, such as Jeans \citep[e.g.][]{Marel94,kleyna01,battaglia08,ReadJustin19,Collins21} or Schwarzschild modeling \citep[e.g.][]{Schwarzschild79,cappellari06,bosch10,Breddels13a,Breddels13b}. 

Despite their strengths, these dynamical models face certain limitations. In Jeans modeling, uncertainties in the stellar velocity anisotropy lead to a well-known degeneracy with the underlying mass profile \citep{Binney82}. Schwarzschild modeling, while more flexible, is sensitive to the quality and completeness of the available data \citep{Kowalczyk17}. Both approaches are also affected by projection effects, which complicate the reconstruction of the galaxy’s three-dimensional structure, and the models typically require significant computational resources and careful treatment of system-specific properties.

In contrast with dynamical modeling, several authors \citep[e.g.,][]{Walker_estim, Wolf_estim, Amorisco_estim, Campbell_estim, Errani_estim} have developed simple formulae to estimate the dynamical mass of galaxies enclosed within specific radii, where they have found velocity anisotropy and/or other factors to introduce minimum uncertainty. These estimators rely on the line-of-sight velocity dispersion of the stellar component of the galaxies and on the half-light radius, and they aim to obtain an unbiased mass estimation that holds for a broad range of galaxy properties and that is minimally affected by projection limitations.

More recently, the combination of machine learning techniques and  hydrodynamical simulations of galaxies has proven to be an effective tool for analyzing complex data and uncovering underlying patterns. These methods have been applied to various astrophysical problems, such as dynamical mass estimation and density profile determination using line-of-sight data. \citet{Ho_2019_clusters} applied a CNN model based on probability distribution functions of positions and velocities of galaxies to estimate the dynamical masses of galaxy clusters. In addition, \citet{nguyen23} developed a graph neural network capable of accurately recovering the DM density profiles of mock spherical dwarf galaxies in dynamical equilibrium. \citet{exposito23} employed 2-d probability distribution functions of projected stellar positions and kinematics as input to a CNN to estimate the inner slopes (150 pc from the center) of DM density profiles in dwarf galaxies. This approach resulted in a model that was able to differentiate between cusps and cores in cosmological hydrodynamical simulations and, when applied to several Milky Way dSphs, was able to recover a similar inner slope as in previous studies \citep{brook15a,ReadJustin19}. \citet{sarrato25} used a similar approach to develop a CNN capable of recovering the full dynamical mass profiles of dispersion-supported galaxies with great success, although encountering a generalization problem when crosstesting the model within different simulation models. \citet{fabio_iocco} used a similar methodology to analyze mock photometry and interferometry images from cosmological hydrodynamical simulations of spiral galaxies, successfully inferring their dynamical mass profiles.

Given the success of such studies, it is natural to ask whether the uncertainty and degeneracies in the inference of DM distributions could be further reduced in galaxies that contain sufficient gas to allow for rotation curve analysis, by simultaneously incorporating the stellar kinematics and the dynamical information provided by the gas component.

In this article, we investigate the potential advantages of using deep learning methods to integrate both gas and stellar velocity fields, a capability that allows for a more complete exploitation of available kinematic information. We extend a probabilistic convolutional neural network, previously developed for stellar kinematic data alone \citep{exposito23,sarrato25}, to incorporate inputs from realistically constructed mock HI observation maps as well.  The model is trained and tested on a large suite of high-resolution cosmological hydrodynamical simulations, with systematic comparisons made between models trained on gas-only, stars-only, and combined kinematic maps. We also evaluate the generalization ability of the network across different simulation suites, in order to assess the model’s robustness to variations in galaxy formation physics.
We find that our model further reduces the scatter in the estimation of dynamical masses compared to the results presented in \citet{sarrato25}, which already demonstrated improved performance over classical mass estimators.

This paper is structured as follows. In Section \ref{sec:dataset}, we describe the galaxy simulations used to train and evaluate the CNN, from the NIHAO \citep{wang15} and AURIGA \citep{Grand17} projects. Section \ref{sec:network} outlines the CNN architectures developed for stellar-only, gas-only, and combined input data, detailing both the network structure and the input/output configurations. In Section \ref{sec:results}, we present the main results of our study. Section \ref{sec:results:nihao} focuses on a comprehensive analysis of model performance using the NIHAO galaxy sample, examining potential biases across different regions of parameter space and comparing the performance of models with varying input types. In Section \ref{sec:results:crosstesting}, we evaluate the model's generalization capability by training and testing across different simulation suites, and we discuss the implications for applying the model to real observational data. We summarise our conclusions in Section \ref{sec:conc}.

\section{Simulation Dataset}
\label{sec:dataset}

We use a comprehensive dataset of realistic galaxy formation simulations to test the ability of a CNN to recover mass profiles. From these simulations, we extract stellar and gas data to serve as input to the network, and compute the corresponding enclosed mass profiles as target outputs. To ensure physical realism, we opt to use cosmological hydrodynamical simulations rather than idealised ones, prioritizing a more representative dataset over a more easily generated but less realistic alternative.

Our goal is to select a dataset that spans the parameter space of galaxy properties as broadly as possible, while still providing a sufficiently large number of galaxies for robust training and evaluation. Simulations from the NIHAO project \citep{wang15} satisfy these criteria. Even so, as demonstrated by \citet{sarrato25}, a model trained and tested on a single simulation model could lead to strong biases, so to assess the generalizability of our model, we additionally incorporate data from the AURIGA dataset \citep{Grand17}.

\subsection{The NIHAO project}
\label{sec:dataset:NIHAO}

The NIHAO project consists in a series of cosmological hydrodynamical zoom-in simulations run with the parameters of \citet{Planck16}: H$_{\text{0}}$ = 100h km s$^{\text{-1}}$ Mpc$^{\text{-1}}$ with h = 0.671, $\Omega_{\text{m}}$ = 0.3175, $ \Omega_{\Lambda}$ = 0.6824, $\Omega_{\text{b}}$ = 0.049 and $\sigma_{\text{8}}$ = 0.8344.

The galaxy formation model incorporates ultraviolet heating, ionization, and metal-line cooling \citep{shen}. Star formation and feedback follow the prescription adopted in the Making Galaxies In a Cosmological Context (MaGICC) simulations \citep{stinson13}, which successfully reproduce galaxy scaling relations over a wide mass range \citep{brook12b}. Star formation occurs in regions exceeding a density threshold of $n_{\rm th} > 10.3~\mathrm{cm}^{-3}$, assuming a \citet{Chabrier03} initial mass function. Stellar energy feedback into the interstellar medium (ISM) is implemented through a combination of blast-wave supernova feedback \citep{stinson06} and early stellar feedback from massive stars. The adopted particle masses and force softenings resolve the mass profile to below 1$\%$ of the virial radius, ensuring that galaxy half-light radii are well resolved.

\subsection{The AURIGA project}
\label{sec:dataset:AURIGA}

The cosmological parameters used in the simulations from the AURIGA project also come from \citet{Planck16}.

The AURIGA galaxy formation model includes magnetohydrodynamics, primordial and metal-line cooling with self-shielding, stellar feedback, and thermal feedback from black holes in both radio and quasar accretion modes. Star formation is implemented according to the Kennicutt–Schmidt law \citep{schmidt59}, adopting a \citet{Chabrier03} initial mass function. The model resolves star-forming regions and feedback processes at high spatial resolution, enabling the formation of thin and thick disc components consistent with those observed in the Milky Way \citep{Grand17}. The resulting galaxies reproduce realistic rotation curves, star formation histories, and structural properties, and follow key empirical relations, including the stellar-to-halo mass relation and the Tully–Fisher relation.

\subsection{Galaxy selection}
\label{sec:dataset:selection}

Our dataset consists of 6158 zoom-in cosmological simulations from the NIHAO project and 956 from the AURIGA project. In addition to the fiducial NIHAO simulations, we include variants featuring different star formation density thresholds $n_{\rm th}$ \citep{nihao20}, simulations incorporating black hole physics \citep{nihao19a}, and high-resolution re-runs of six NIHAO halos \citep{nihao19b}.

We select all galaxies, satellites included, with stellar masses in the range of M$_{*} = 10^{5.5}$–$10^{11}$ M$_{\odot}$ containing at least 100 stellar and gas particles and a high-resolution particle mass fraction exceeding 95\%. To further enhance dataset quality, we visually inspect all selected galaxies and manually exclude ongoing mergers and severely disrupted systems. We refer to this dataset as the \textbf{full set}.

We also construct a limited set with the further constrains of cold gas masses of M$_{\rm cold} > 10^{7.5}$ M$_{\odot}$ and M$_{\rm cold}$/M$_{*} > 0.1$, where we use the mass of HI as proxy for the cold gas, reducing the number of galaxies to 5001 from NIHAO and 907 from AURIGA. This set ensures that the galaxies have a relevant amount of cold gas. We refer to this dataset as the \textbf{cold gas set}.

\section{Neural Network}
\label{sec:network}

In this work, we tackle the problem of inferring the dark matter content underlying the kinematic, dynamical, and photometric properties of galaxies. This constitutes a complex pattern recognition task, well-suited to modern deep learning approaches. In particular, convolutional neural networks (CNNs) are an effective tool for capturing and modeling the structured information present in such multidimensional datasets.

We use the python's package \textsc{tensorflow} \citep{tensorflow} for constructing a CNN based on the one used in \citet{exposito23} and \citet{sarrato25}.

\subsection{Input data}
\label{sec:network:input}

To construct the inputs of our network, we projected all our galaxies on the plane of the sky in 12 different orientations with increasing inclination. For each of these projections we constructed two maps with star information (Sec. \ref{sec:network:input:star}) and three maps with cold gas information (Sec. \ref{sec:network:input:gas}).

\subsubsection{Star data}
\label{sec:network:input:star}

Typically, the number of stars for which spectroscopic data is available for LG dwarf galaxies is in the order of hundreds or thousands, while the number of star particles available in our simulated galaxies range from a few hundred to several million, with a mean number of about $10^5$ stellar particles in each galaxy.

Therefore, in order to simulate an observational sample of stars, we divided each simulated galaxy's complete sample of stars into several subsets, each made of randomly selected stars, and we use just one of those subsets for each projection of our dataset. The number of stars within each subset of a given galaxy is dependent on the total number of star particles in the simulation,  with an upper limit of $10^4$ stars and a lower limit of 100 stars.

These projected stars are defined by their position ($x_{\rm proj}$, $y_{\rm proj}$) and their line-of-sight velocity $v_{\rm LOS}$; we use this information to construct the images with star information that function as inputs of our CNN. The generated images images are continuous 2D probability density functions (PDFs) of the distribution of stars in projected phase spaces, constructed with bivariate kernel density estimations (KDEs). The mapping generated with KDEs allows us to encapsulate the features of the original discrete distributions in the same form even if each galaxy subset is represented by a different number of stars. We construct two maps:

\begin{itemize}
    \item A PDF sampled at 64x64 points with the distribution of stars in \{x,y\} phase space, between -0.5 $R_{\rm hl}$ and 0.5 $R_{\rm hl}$ in each coordinate, in the reference system where (x,y) = (0,0) is the center of the galaxy. $R_{\rm hl}$ is the projected half-stellar count radius of the galaxy, which we use as a proxy for the half-light radius.
    \item A PDF sampled at 64x64 points with the distribution of stars in  \{$\hat{R}_{\rm proj},\hat{v}_{\rm LOS}$\} phase space, where $\hat{R}_{\rm proj} = \sqrt{x^2 + y^2}/R_{\rm hl}$ is the radial position normalized by the half-light radius and $\hat{v}_{\rm LOS} = v_{\rm LOS}/P_{98\%}$ is the line-of-sight velocity normalized by the 98\% percentile of the absolute value of $v_{\rm LOS}$ of all stars of the sample. $\hat{R}_{\rm proj}$ ranges from 0 to 1, and $\hat{v}_{\rm LOS}$ ranges from -1 to 1. 
\end{itemize}

Note that the adopted spatial limits effectively exclude stellar data beyond these regions. During the testing phase, multiple boundary conditions and normalization strategies were evaluated. The limits used in this study correspond to those that yielded the best overall performance on the current dataset. A plausible explanation is that including such data in the input probability distribution functions (PDFs) may dilute the more informative signal from stars located nearer to the galaxy center, where the gravitational potential is most tightly constrained.

\begin{figure}[ht]
    \centering
    \includegraphics[width = \columnwidth]{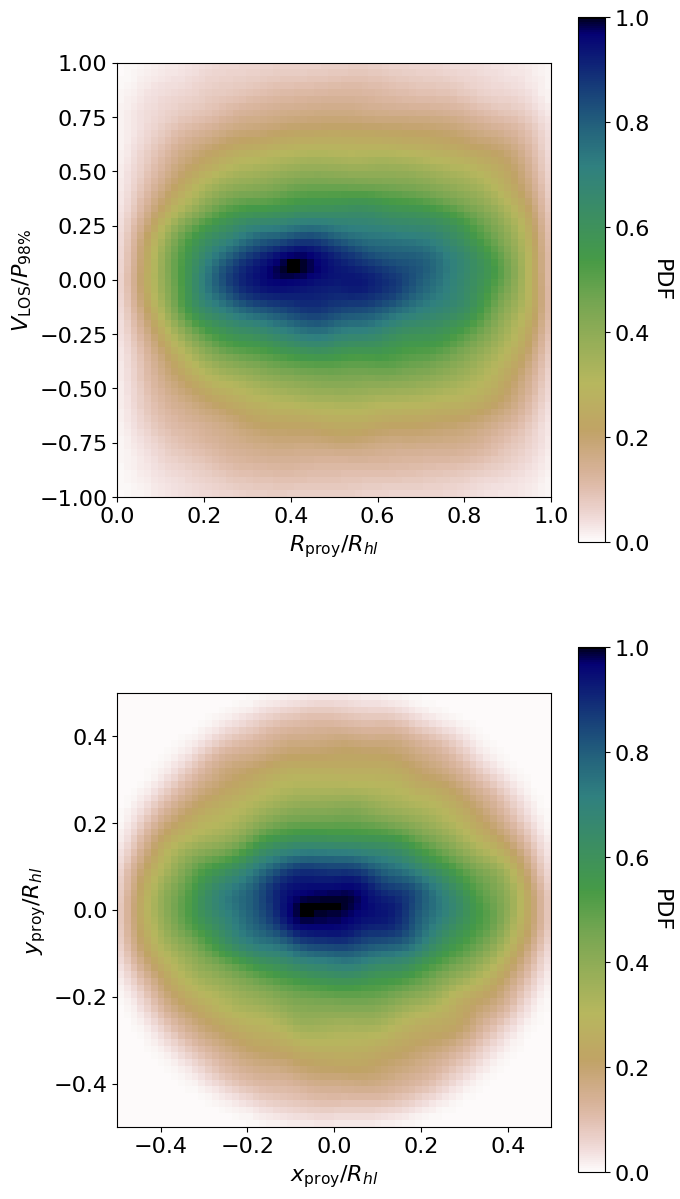}
    \caption{Star information input maps for a single edge-on galaxy from the NIHAO sample of M$_* = 10.68$ M$_\odot$. Top: PDF in the \{$\hat{R}_{\rm proj},\hat{v}_{\rm LOS}$\} phase space. Bottom: PDF in the \{x,y\} phase space. Both obtained following the procedure described in section \ref{sec:network:input:star}.}
    \label{fig:star_maps}
\end{figure}

In Fig. \ref{fig:star_maps} we show an example of the star maps for an edge-on galaxy from NIHAO.

\subsubsection{Gas data}
\label{sec:network:input:gas}

We generate mock HI observations using the \textsc{MARTINI} code, described in \citet{oman19}. This tool enables the production of synthetic, spatially resolved HI line data cubes directly from hydrodynamical simulation snapshots. It offers a comprehensive suite of features, including spectral modeling and the incorporation of observational effects such as noise contamination and beam convolution.

For the same galaxy projections used to create the star maps as explained in \ref{sec:network:input:star}, we use \textsc{MARTINI} to create a spectral data cube of 64 channels with a spectral resolution of 5 km/s. The input for \textsc{MARTINI} consists of the position and average velocities of the gas particles, their HI component mass, their temperature and their softening length, along with the distance to the observer, fixed at 1 Mpc as a typical distance for LG galaxies \citep{walker03}. We set the space covered by the HI line data cubes from the center of the galaxy as double the radius where the HI superficial density drops below 1 M$_\odot$ pc$^{-2}$, R$_{\Sigma_{\rm HI} < 1 {\rm M}_\odot {\rm pc}^{-2}}$.

The resulting data cubes comprise 64 spectral channels; to reduce dimensionality while preserving essential information, we extracted the first three statistical moments of the HI emission: 

\begin{itemize}
    \item A 64x64 grid with a side length of 4 R$_{\Sigma_{\rm HI} < 1 {\rm M}_\odot {\rm pc}^{-2}}$, with the line-of-sight integrated intensity.
    \item A 64x64 grid with the average line-of-sight velocity covering the same physical space.
    \item A 64x64 grid with the line-of-sight velocity dispersion covering the same physical space.
\end{itemize}

\begin{figure*}[ht]
    \centering
    \includegraphics[width = 420pt]{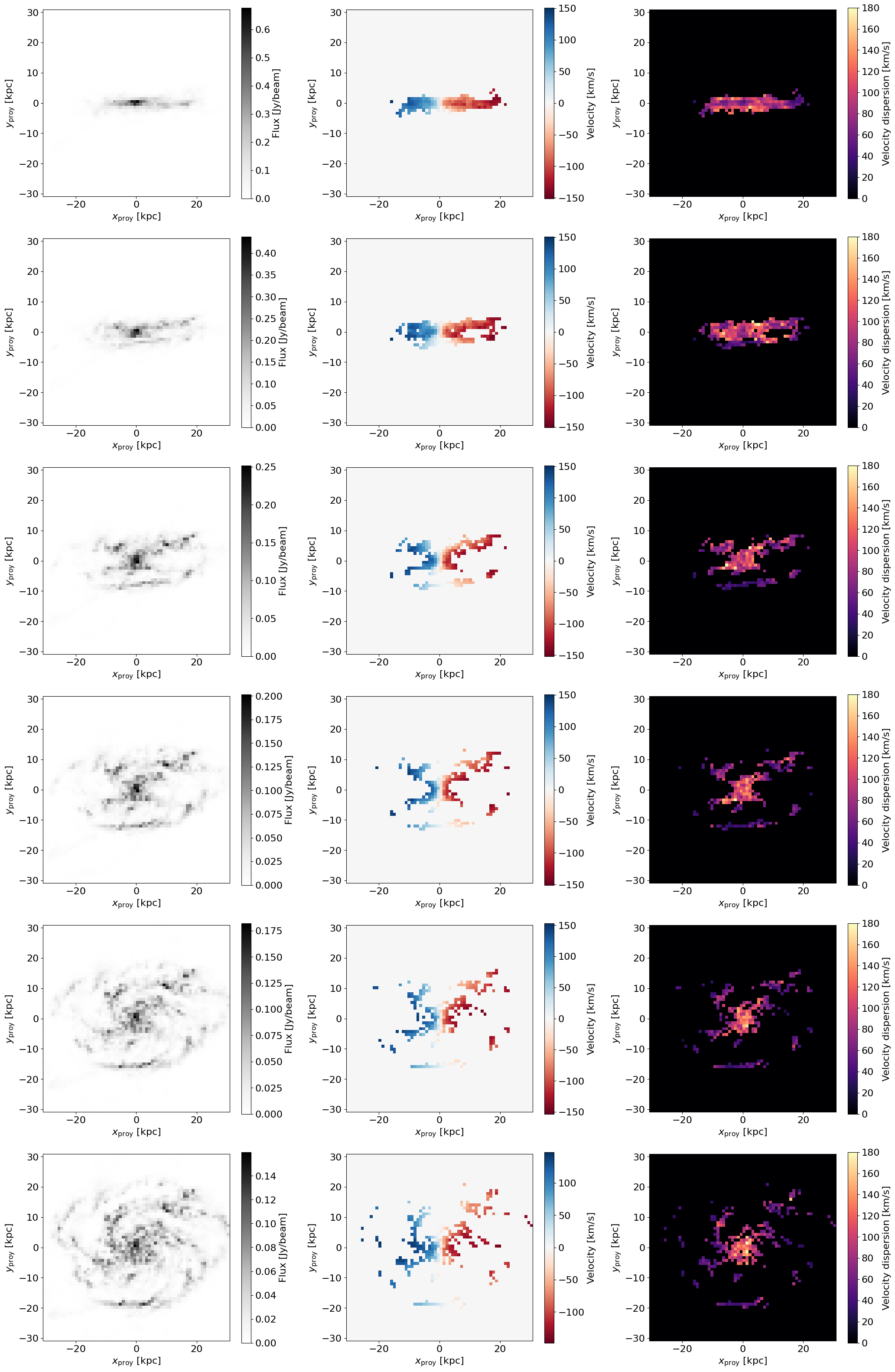}
    \caption{Gas information input maps for a single galaxy from the NIHAO sample of M$_* = 10.68$ M$_\odot$ at increasing inclinations from edge-on to 60 degrees. From left to right: HI intensity, average line-of-sight velocity and velocity dispersion maps, obtained following the procedure described in section \ref{sec:network:input:gas}.}
    \label{fig:gas_maps}
\end{figure*}

To ensure a sufficient signal-to-noise ratio for the first and second moments, we applied a mask to exclude regions with a column density below 10$^{19.5}$ atoms cm$^{-2}$, as recommended by \citet{oman19}. In Fig. \ref{fig:gas_maps} we show an example of the gas maps for different inclinations of a NIHAO galaxy.

\subsection{Architecture}
\label{sec:network:architecture}

\begin{figure*}[ht]
    \centering
    \includegraphics[width = \textwidth]{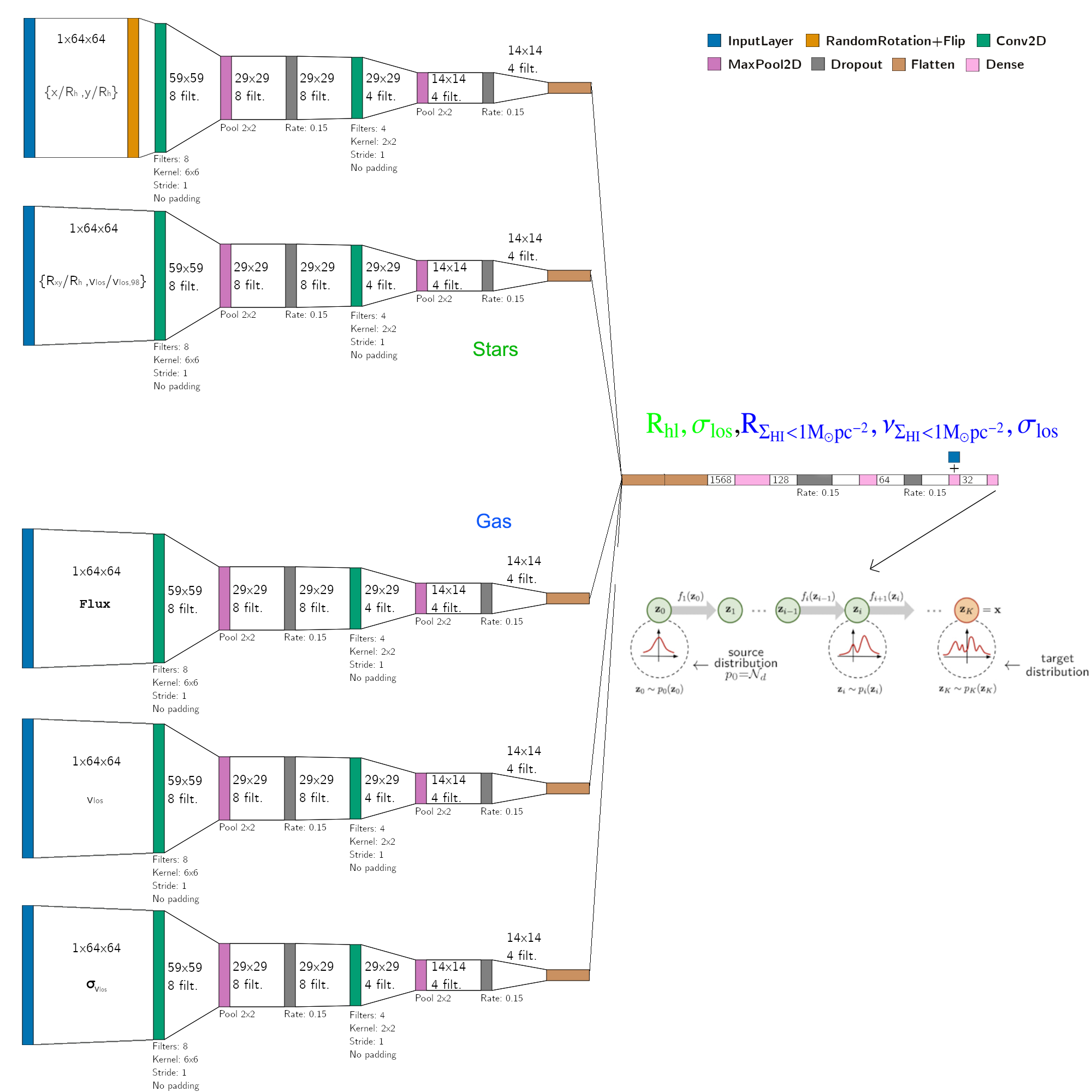}
    \caption{Scheme of our CNN architecture. Five parallel branches (two for star information, three for gas information) pass through several convolutional and pooling layers independly, further reducing the dimensionality, and then they merge for a final sequence of dense and pooling layers, producing a N parameter output. After training the CNN, the 32 neurons of penultimate layer are used as inputs to train a normalising flow model. The flow model learns a series of transformations to a N-dimensional gaussian PDF, which are conditioned on the inputs, and outputs a posterior N-dimensional joint PDF. The final output represents the value estimated for the dynamical mass of the galaxy enclosed within N different radii.}
    \label{fig:architecture}
\end{figure*}

The architecture of our neural network is shown in Fig. \ref{fig:architecture}. The CNN is organized into five parallel branches, each corresponding to one of the input map types introduced in Secs. \ref{sec:network:input:star} and \ref{sec:network:input:gas}. This design allows the network to learn feature representations that are specific to the stellar and gas components before they are combined. Within each branch, the input is processed by two convolutional blocks, each consisting of a convolutional layer followed by max-pooling, which extracts spatial features and compresses the information into progressively higher-level representations. To reduce overfitting and improve the ability to generalize across galaxies with different morphologies and inclinations, dropout layers are applied after these blocks. The outputs of the five branches are then flattened into one-dimensional vectors and concatenated into a single feature representation. This combined vector is passed through a sequence of fully connected layers that capture correlations across the stellar and gas channels while gradually reducing dimensionality. The final output of the network consists of $N$ values corresponding to the predicted enclosed mass profile. In this study, we use $N=10$, with points distributed between $0.6 \ R_{\rm hl}$ and $2.4 \ R_{\rm hl}$. Dropout layers are also included between selected dense layers to improve training stability.

To provide a scale to the normalised input data, we expand this representation with a set of global galaxy parameters, concatenated to the output of the first dropout layer after the concatenation. These include the projected half-light radius $R_{\rm hl}$ and the line-of-sight velocity dispersion $\sigma_{\rm v}$. For the gas component, we add both structural and dynamical properties: the radius at which the HI surface density falls below $1 \ M_\odot \ {\rm pc}^{-2}$ ($R_{\Sigma_{\rm HI} < 1 {\rm M}_\odot {\rm pc}^{-2}}$), the rotational velocity of the cold gas at this radius ($v_{\rm cold}$), and the mean cold gas velocity dispersion within the same region ($\sigma_{\rm cold}$).

A central feature of this design is its modularity. By selectively activating input branches and global parameters, the model can be trained in several configurations: using only stellar data, only gas data, or both combined. This flexibility is crucial for our broader goal of testing how different tracers contribute to the accuracy and robustness of dynamical mass inferences.

To produce reliable uncertainty estimates, we extend the CNN with a normalizing flow. The 32-dimensional feature vector from the penultimate dense layer serves as input to a normalizing flow model \citep{normflows}, implemented with the ltu-ili Python package \citep[Learning the Universe Implicit Likelihood Inference;][]{ltuili}. We adopt a Masked Autoregressive Flow (MAF) \citep{maf}, which learns a set of invertible transformations conditioned on the CNN features, mapping a simple Gaussian distribution into the posterior over enclosed masses. The resulting model outputs an $N$-dimensional joint probability density function (PDF) that captures correlations between the mass estimates at different radii. This probabilistic approach provides a principled way to assess uncertainties, ensuring that the predictions are not only accurate but also interpretable. In the context of our study, this capability is essential for identifying systematic biases, evaluating generalization across simulation suites, and ultimately guiding the application of the model to real observational data.

\section{Results}
\label{sec:results}

\subsection{NIHAO results}
\label{sec:results:nihao}

We define three different models for our network based on the number of channels and the given information.

\begin{itemize}
    \item \textbf{Gas model:} The inputs are the three gas channels described in Sec. \ref{sec:network:architecture}, the radius where the HI superficial density drops below 1 M$_\odot$ pc$^{-2}$ (R$_{\Sigma_{\rm HI} < 1 {\rm M}_\odot {\rm pc}^{-2}}$), the velocity of the gas at that radius and the mean velocity dispersion of the gas inside that radius.
    \item \textbf{Star model:} The inputs are the two star channels described in Sec. \ref{sec:network:architecture}, the projected half-light radius and the mean velocity dispersion of the stars.
    \item \textbf{StarGas model:} The inputs are all of the above: the five channels described in Sec. \ref{sec:network:architecture}, the value of R$_{\Sigma_{\rm HI} < 1 {\rm M}_\odot {\rm pc}^{-2}}$, the velocity of the gas at that radius and the mean velocity dispersion of the gas inside that radius, the projected half-light radius and the mean velocity dispersion of the stars.
\end{itemize}

\begin{figure}[ht]
    \centering
    \includegraphics[width = \columnwidth]{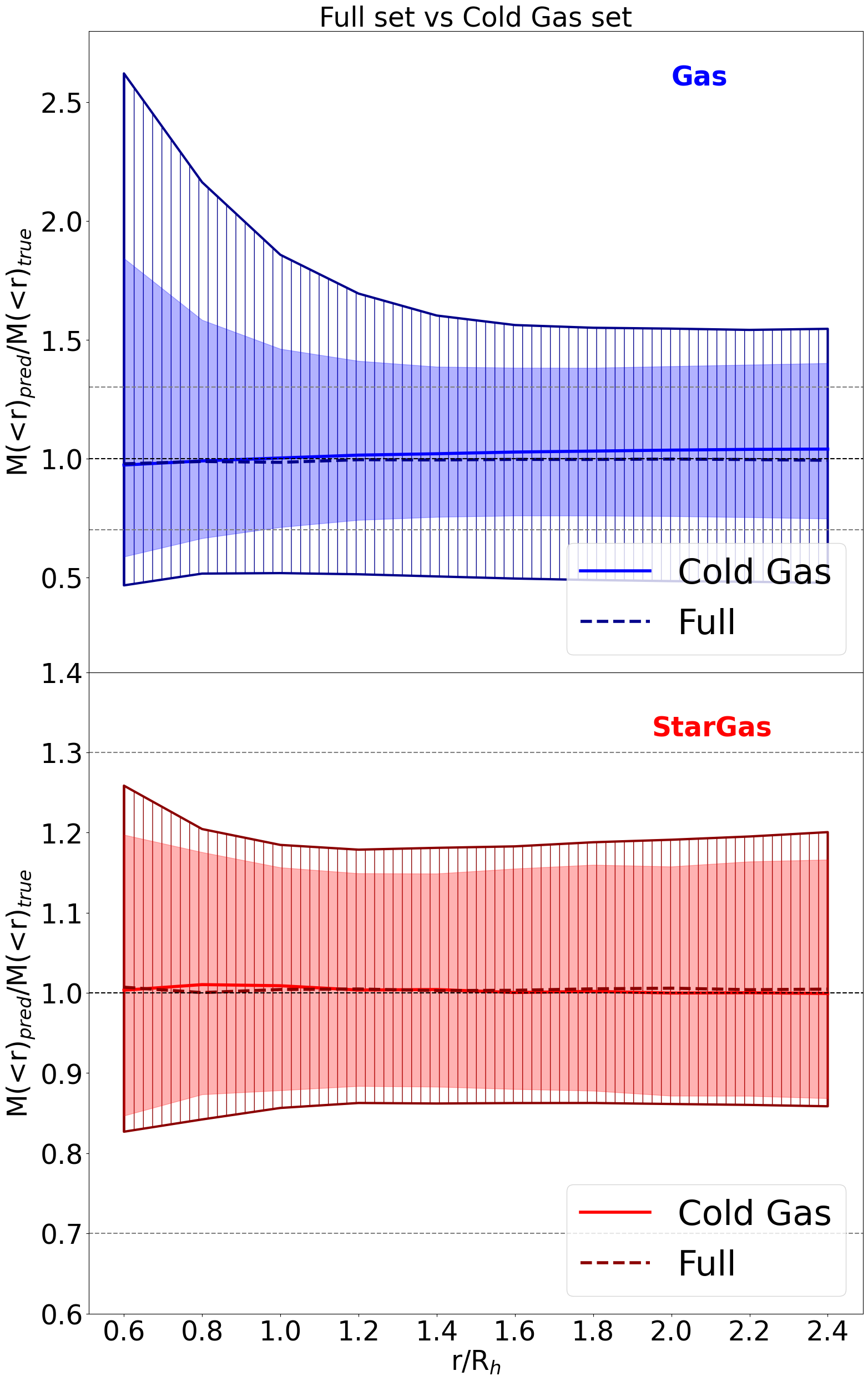}
    \caption{Ratio between the mass predicted by the neural network models and the real mass enclosed within different radii of the galaxies in the test sets of NIHAO galaxy projections. Top: Results for the Gas model. Bottom: Results for the StarGas model.} 
    \label{fig:Full_vs_ColdGas}
\end{figure}

In Fig. \ref{fig:Full_vs_ColdGas} we compare the results for the full set (M$_{*} = 10^{5.5}$–$10^{11}$ M$_{\odot}$ containing at least 100 stellar and gas particles and a high-resolution particle mass fraction exceeding 95\%) and the cold gas set (same restrictions, plus cold gas masses of M$_{\rm cold} > 10^{7.5}$ M$_{\odot}$ and M$_{\rm cold}$/M$_{*}$ > 0.1), using both the Gas model and the StarGas model. 

\clearpage

Only 18.8\% of the galaxies from the full set don't meet the cold gas set criteria, but their presence in the dataset increases the dispersion on the enclosed mass prediction significantly, being 1.7 times that of the cold gas set, due to to the lack of cold gas in these galaxies. On the other hand, the StarGas model suffers only a slight increase in homogeneous dispersion over the whole range of distances when using the full set instead of the cold gas set, which shows that the full model is relatively stable in the presence of galaxies without gas information in the dataset and is able to make use of the stellar information provided in those cases even in a global training.

As we are interested in studying the performance of the model when making use of a dual source of information, we will use the cold gas set throughout the rest of the analysis.

\begin{figure}[ht]
    \centering
    \includegraphics[width = \columnwidth]{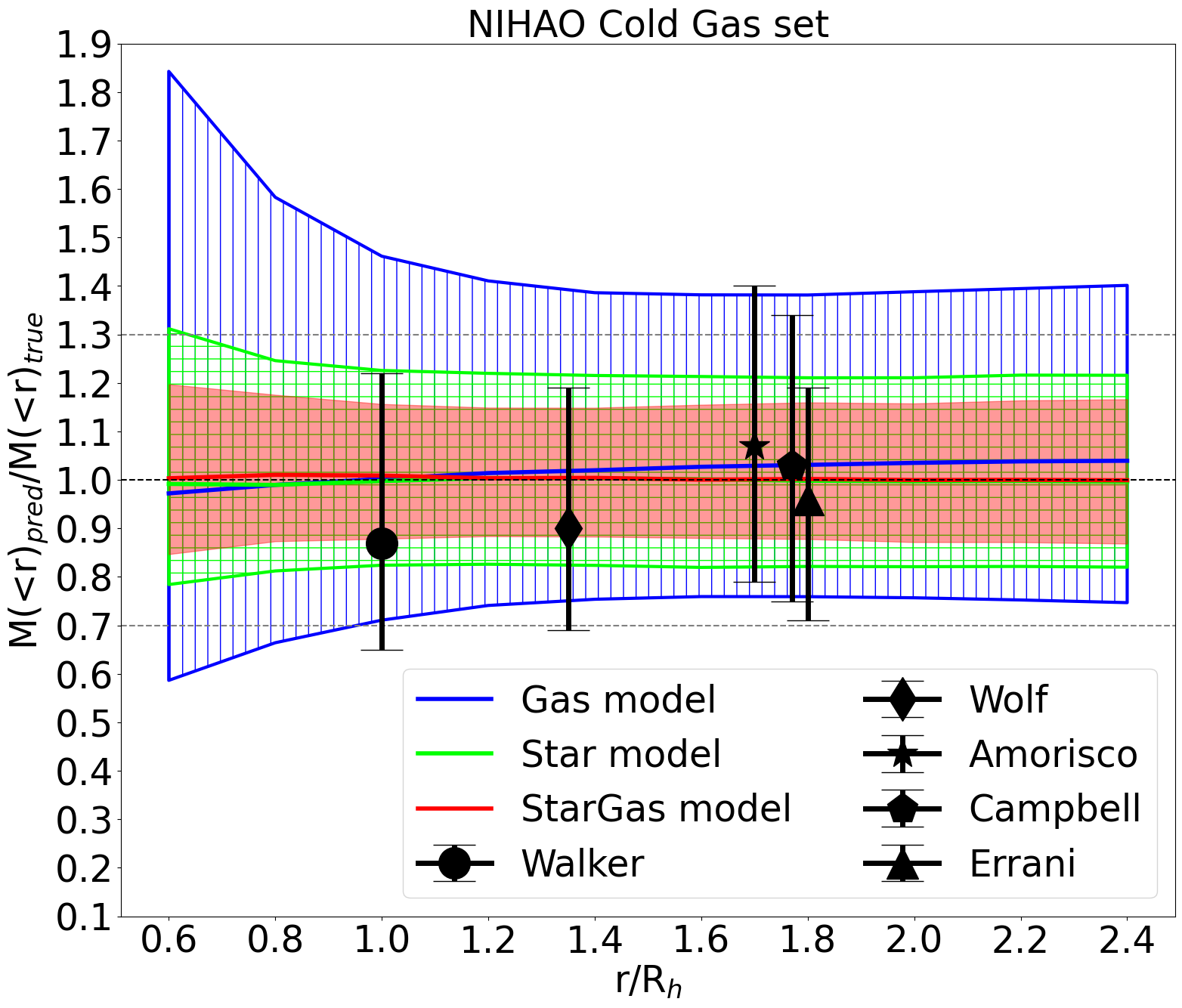}
    \caption{Ratio between the mass predicted by the neural network models and the real mass enclosed within different radii of the galaxies in the test sets of NIHAO galaxy projections. The colored lines show the median ratio using the mass estimated by the CNN for the training set, while the shadowed regions indicate the 1-$\sigma$. Blue: Results when using the three gas channels described in Sec. \ref{sec:network:architecture}, the radius where the cold gas superficial density falls below 1 M$_\odot$ pc$^{-2}$ and the mean velocity dispersion inside that radius. Green: Results when using the two star channels described in Sec. \ref{sec:network:architecture}, the projected half-light radius and the mean velocity dispersion of the stars. Red: Results when using all of the above. The predicted to true enclosed mass ratios resulting from applying literature mass estimators to NIHAO galaxies are shown as black symbols with 1-$\sigma$ errorbars.}
    \label{fig:MvsR}
\end{figure}

In Fig. \ref{fig:MvsR} we show the accuracy of the model trained on NIHAO galaxies at all radii for the testing set and using the three models. Numerical results at 0.6R$_{\rm hl}$, R$_{\rm hl}$ and 2.4R$_{\rm hl}$ are shown in Table \ref{tab:MvsR}. The bias in the recovery of the mass at different radii is negligible in all three cases, with a maximum error for any of the radii of 0.03, although the Gas model presents a tendency to underestimate enclosed mass at lower radii and overestimate it at higher radii. On the other hand, in comparison to the Star model, the 1-$\sigma$ uncertainty is reduced by a factor $\sim 1.5$ throughout the profile when using gas information along with the star information, with a particularly significant improvement in the innermost part of the galaxy. The combination of gas and star information allows the model to improve its prediction at all studied radii. 

The Gas model, only by itself, performs significantly worse than the Star and StarGas models at all radii, with dispersions of $\sim 2.2$ times those of the StarGas model in the outer part of the galaxy and above $\sim 3.5$ in the inner side, making the model unsuitable to be used just with cold gas information from the galaxies. These results point to a possible decoupling between gas morphology and gravitational potential in NIHAO dwarf galaxies, likely driven by strong feedback-related effects.

\begin{table}
\resizebox{\columnwidth}{!}{%
\begin{tabular}{|c|c|c|c|} 
\hline
\multicolumn{4}{|c|}{M(<r)$_{\rm pred}$/M(<r)$_{\rm true}$} \\
\hline
\hline
NN model & 0.6R$_{\rm hl}$ & R$_{\rm hl}$ & 2.4R$_{\rm hl}$\\
\hline
Star & $0.99^{+0.32}_{-0.21}$ & $1.00^{+0.23}_{-0.17}$ & $1.00^{+0.22}_{-0.18}$\\
\hline
Gas & $0.97^{+0.87}_{-0.39}$ & $1.00^{+0.46}_{-0.29}$ & $1.03^{+0.36}_{-0.29}$\\
\hline
StarGas & $1.00^{+0.19}_{-0.16}$ & $1.00^{+0.14}_{-0.13}$ & $1.00^{+0.17}_{-0.13}$\\
\hline
\end{tabular}%
}
\caption{Mean value and 1$\sigma$ scatter of the predicted to true enclosed mass ratio M(<$r$)$_{\rm pred}$/M(<$r$)$_{\rm true}$ for the Star, Gas and StarGas model, at 0.6R$_{\rm hl}$, R$_{\rm hl}$ and 2.4R$_{\rm hl}$. Values from Fig. \ref{fig:MvsR}.}
\label{tab:MvsR}
\end{table}

We also compare the results with several mass estimators \citep{Walker_estim, Wolf_estim, Amorisco_estim, Campbell_estim, Errani_estim}. The StarGas model shows less error in the mass recovery for all cases while at the same time maintaining a negligible bias.

\begin{figure}[ht]
    \centering
    \includegraphics[width = \columnwidth]{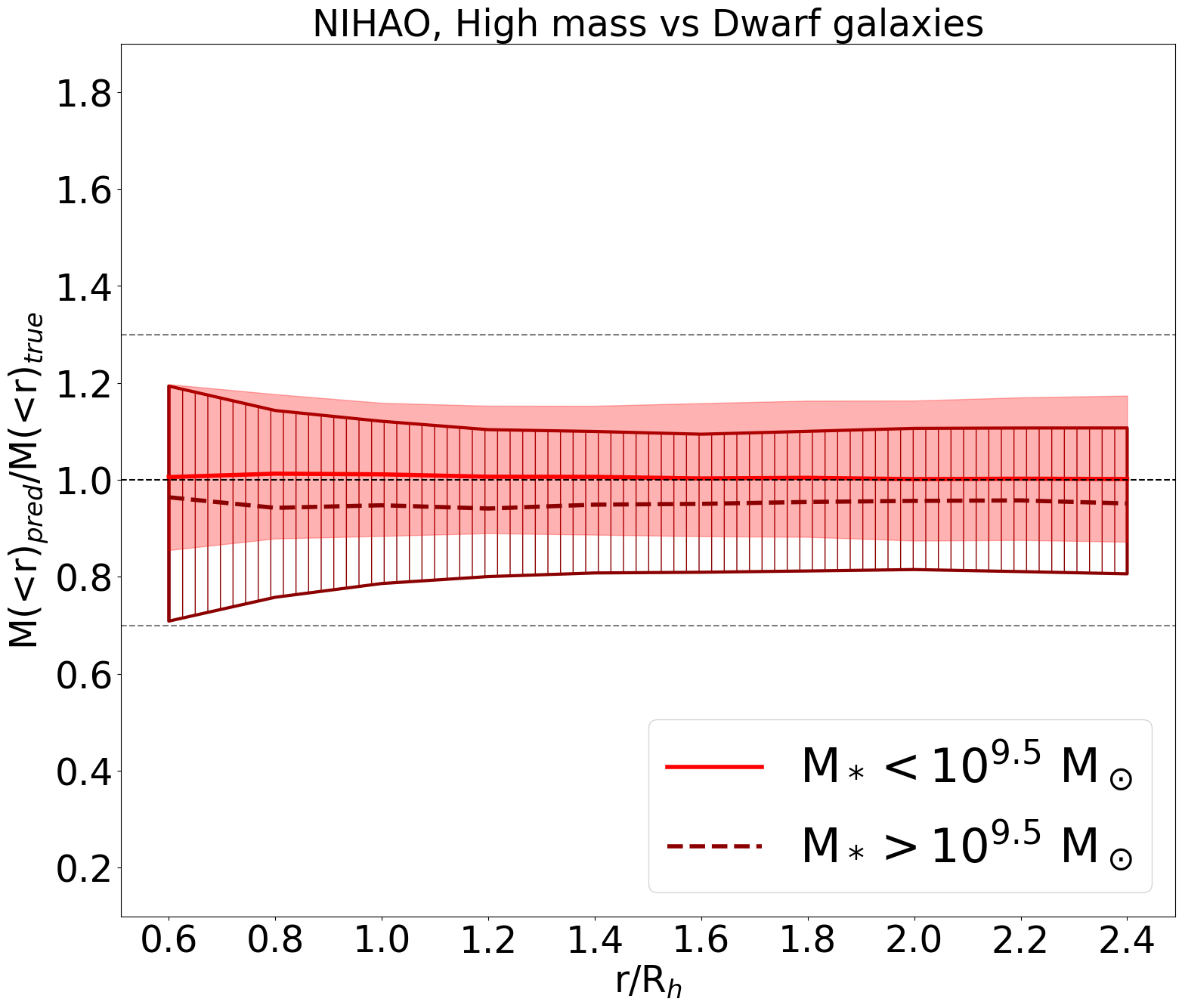}
    \caption{Ratio between the mass predicted by the neural network models and the real mass enclosed within different radii of the galaxies in the test sets of NIHAO galaxy projections. Results for the StarGas model applied to two subsets of galaxies with M$_{\rm *} < 10^{9.5}$ M$_\odot$ and M$_{\rm *} > 10^{9.5}$ M$_\odot$.} 
    \label{fig:High_vs_Dwarf}
\end{figure}

We next test if the model, trained over such a wide mass range, can be applied equally well to low and high mass galaxies. In Fig. \ref{fig:High_vs_Dwarf} we show the accuracy of the StarGas model for a testing set of only dwarf galaxies (M$_{\rm *} < 10^{9.5}$ M$_\odot$) and high mass galaxies (M$_{\rm *} > 10^{9.5}$ M$_\odot$). The StarGas model show a clear bias towards underestimating the masses of high mass galaxies, up to predicting less than $\sim$7$\%$ of the total mass in the average prediction of the entire high mass subset.

This bias is most likely a direct consequence of an overrepresentation of dwarf galaxies in the training set; from the total of 4959 galaxies in the training dataset, only 328 have a stellar mass greater than $10^{9.5}$ M$_\odot$ due to the large number of satellites in the dataset. This may have led to an improvement in the optimization of the complete training by tending to reduce the overall estimated mass, even if that increased the error in the massive galaxy part of the set.

\begin{figure}[ht]
    \centering
    \includegraphics[width = \columnwidth]{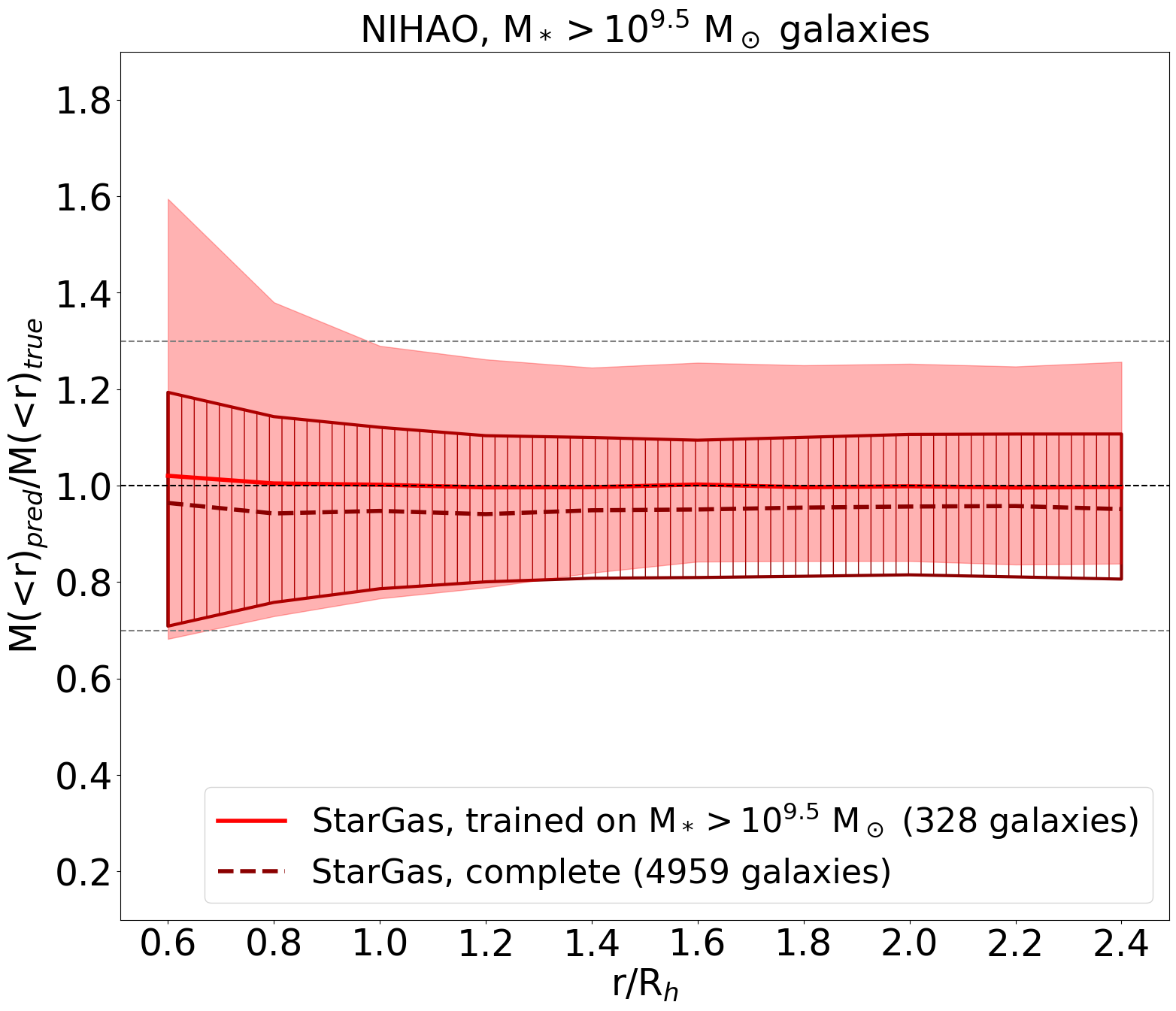}
    \caption{Ratio between the mass predicted by the StarGas model and the real mass enclosed within different radii of the galaxies in the test sets of NIHAO galaxy projections. Results applied to a subset of galaxies with M$_{\rm *} > 10^{9.5}$ M$_\odot$ for the main StarGas model and for a modified StarGas model trained only with a subset of 328 high mass galaxies with M$_{\rm *} > 10^{9.5}$ M$_\odot$.} 
    \label{fig:High_vs_Dwarf_2}
\end{figure}

The effect of training a model only with high mass galaxies can be seen in Fig. \ref{fig:High_vs_Dwarf_2}. When training only on high mass galaxies the bias towards underestimating their mass is eliminated. On the other hand, the dispersion multiplies by a factor of 3 in comparison to the dispersion presented in the model trained with all the galaxies. This shows that a greater number of galaxies, even if they are dwarf galaxies, gives relevant information to the model in regard to inferring mass profiles for high mass galaxies, even if it also generated a bias. The way of dealing with this effect will be explored in future work.

\begin{figure*}[ht]
    \centering
    \includegraphics[width = \textwidth]{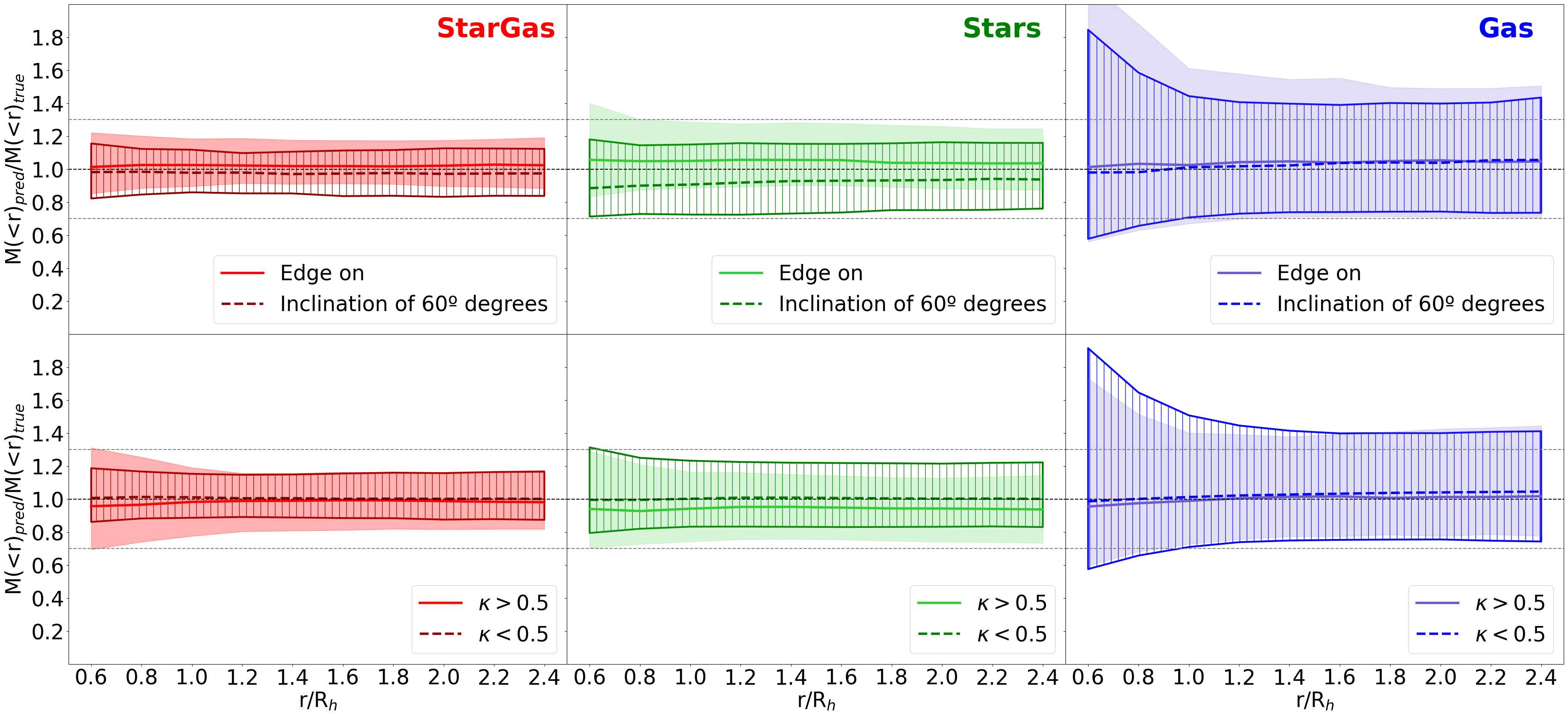}
    \caption{Ratio between the mass predicted by the neural network models and the real mass enclosed within different radii of the galaxies in the test sets of NIHAO galaxy projections. Results for two subsets consisting of rotation supported galaxies with $\kappa > 0.5$ and other consistent of dispersion supported galaxies with $\kappa < 0.5$ and for two subsets consistent of edge-on galaxies and galaxies at an inclination of 60\textdegree degrees.} 
    \label{fig:Inc_and_RotDisp}
\end{figure*}

We also investigate the performance of the models on galaxies with differing kinematic support, specifically comparing rotation-supported and dispersion-supported systems. Dispersion-supported galaxies are identified using the criterion $\kappa_{\textrm{co}} < 0.5$. The parameter $\kappa_{\textrm{co}}$, introduced in \citet{salvador17}, quantifies the fraction of stellar kinetic energy invested in ordered rotation and is defined as:

\begin{equation}
    \label{eq:kappa}
    \kappa_{\textrm{co}} = \frac{1}{K_{\textrm{s}}} \sum_{i}^{r < \textrm{30 kpc} \ ; \ j_{\textrm{z,i}}>0} \frac{m_{\textrm{i}}}{2} \left( \frac{j_{\parallel,\textrm{i}}}{R_{\perp,\textrm{i}}} \right)^{2},
\end{equation}

where $K_{\textrm{s}}$ is the total kinetic energy of the stellar component, and the sum is taken over all star particles within 30 kpc of the galaxy’s center that have positive specific angular momentum $j_{\parallel}$, aligned with the total stellar angular momentum vector, $m_{\textrm{i}}$ is the mass of the $i$-th stellar particle, and $R_{\perp,\textrm{i}}$ denotes its distance from the galaxy's stellar rotation axis.

We show the results in Fig. \ref{fig:Inc_and_RotDisp}. The Star model works significantly worse with rotation-supported galaxies than with dispersion-supported galaxies and the StarGas model also shows a clear increase in error in mass recovery for rotation-supported galaxies. A similar bias to what we encountered with high mass galaxies is present with the Star model, but it is mostly eliminated on the StarGas model. The Gas model shows much greater dispersion, but no bias, and a better performance on rotation-supported galaxies than on dispersion-supported ones, as expected. It is therefore reasonable to conclude that the information provided by gas is allowing us to correct the bias found when using only stellar information, while reducing dispersion even though it is much greater when using gas alone.

We finally study the effect of inclination in the model predictions. In Fig. \ref{fig:Inc_and_RotDisp}, we compare the performance of the three models on a subset of the testing dataset consisting exclusively of edge-on galaxies and galaxies inclined at $60^\circ$ with respect to the line of sight.

The Star model exhibits a systematic bias with respect to galaxy inclination. For edge-on galaxies, the enclosed mass is consistently overestimated across all radii, with a mean ratio of M(<r)$_{\rm pred}$/M(<r)$_{\rm true} = 1.07$. In contrast, for galaxies inclined at $60^\circ$, the model underestimates the mass, with M(<r)$_{\rm pred}$/M(<r)$_{\rm true} = 0.89$ at small radii and $0.96$ at larger radii. These trends indicate that the Star model lacks the capacity to infer galaxy inclination and appropriately correct its mass predictions. Instead, it interprets the inclination-induced suppression of projected stellar velocities as a signature of lower mass. During training, the model compensates for this effect by adjusting its predictions to minimize the overall error across the full dataset, where galaxies with all inclinations are present. As a result, inclination-dependent biases persist in the model's output. The Star model is, thus, strongly subject to projection effects, as happens with dynamical modeling.

The Gas model, instead, does not exhibit a significant inclination-dependent bias; however, it shows increased scatter in its predictions for edge-on galaxies. This may be due to a stronger degeneracy between edge-on disc galaxies and spheroidal systems observed at various inclinations. In contrast, non-spheroidal galaxies at intermediate inclinations may be more easily distinguishable from spheroidal ones, enabling the gas-related input channels to better infer the underlying gravitational potential.

The StarGas model achieves significantly lower dispersion than either the Star or Gas models and displays only a minor bias, likely inherited from the stellar input channels. Nevertheless, the mean residuals of the predicted to true enclosed mass ratio remain below $0.05$ at all radii for both inclination subsets. This again highlights the model’s ability to partially correct the individual limitations of using stellar or gas information alone by leveraging their combined input.

\subsection{Crosstesting between different simulation datasets}
\label{sec:results:crosstesting}

\begin{table}
\resizebox{\columnwidth}{!}{%
\begin{tabular}{|c|c|c|c|c|} 
\hline
\multicolumn{5}{|c|}{M(<r)$_{\rm pred}$/M(<r)$_{\rm true}$, StarGas model} \\
\hline
\hline
  Trained & Tested & 0.6R$_{\rm hl}$ & R$_{\rm hl}$ & 2.4R$_{\rm hl}$\\
\hline
\multirow{2}{*}{NIHAO} & NIHAO   & $0.99^{+0.24}_{-0.16}$ & $1.00^{+0.16}_{-0.13}$ & $1.01^{+0.17}_{-0.14}$\\
\cline{2-5}
& AURIGA  & $0.94^{+0.37}_{-0.22}$ & $0.95^{+0.32}_{-0.20}$ & $0.96^{+0.26}_{-0.17}$\\
\hline
\multirow{2}{*}{AURIGA} & NIHAO    & $0.93^{+1.07}_{-0.32}$ & $0.97^{+0.57}_{-0.25}$ & $1.03^{+0.36}_{-0.40}$\\
\cline{2-5}
& AURIGA  & $1.01^{+0.26}_{-0.22}$ & $1.01^{+0.21}_{-0.20}$ & $1.01^{+0.14}_{-0.15}$\\
\hline
\multirow{2}{*}{BOTH} & NIHAO  & $1.00^{+0.23}_{-0.14}$ & $1.02^{+0.18}_{-0.16}$ & $1.00^{+0.22}_{-0.12}$\\
\cline{2-5}
& AURIGA  & $0.98^{+0.28}_{-0.22}$ & $0.98^{+0.24}_{-0.20}$ & $0.99^{+0.20}_{-0.16}$\\
\hline
\end{tabular}%
}
\caption{Mean value and 1$\sigma$ scatter of predicted to true enclosed mass ratio M(<$r$)$_{\rm pred}$/M(<$r$)$_{\rm true}$ for various training and testing combinations across NIHAO and AURIGA simulation suites using the StarGas model, at 0.6R$_{\rm hl}$, R$_{\rm hl}$ and 2.4R$_{\rm hl}$. Values from Fig. \ref{fig:CrossTesting}.}
\label{tab:CrossTesting}
\end{table}

\begin{figure*}[ht]
    \centering
    \includegraphics[width = \textwidth]{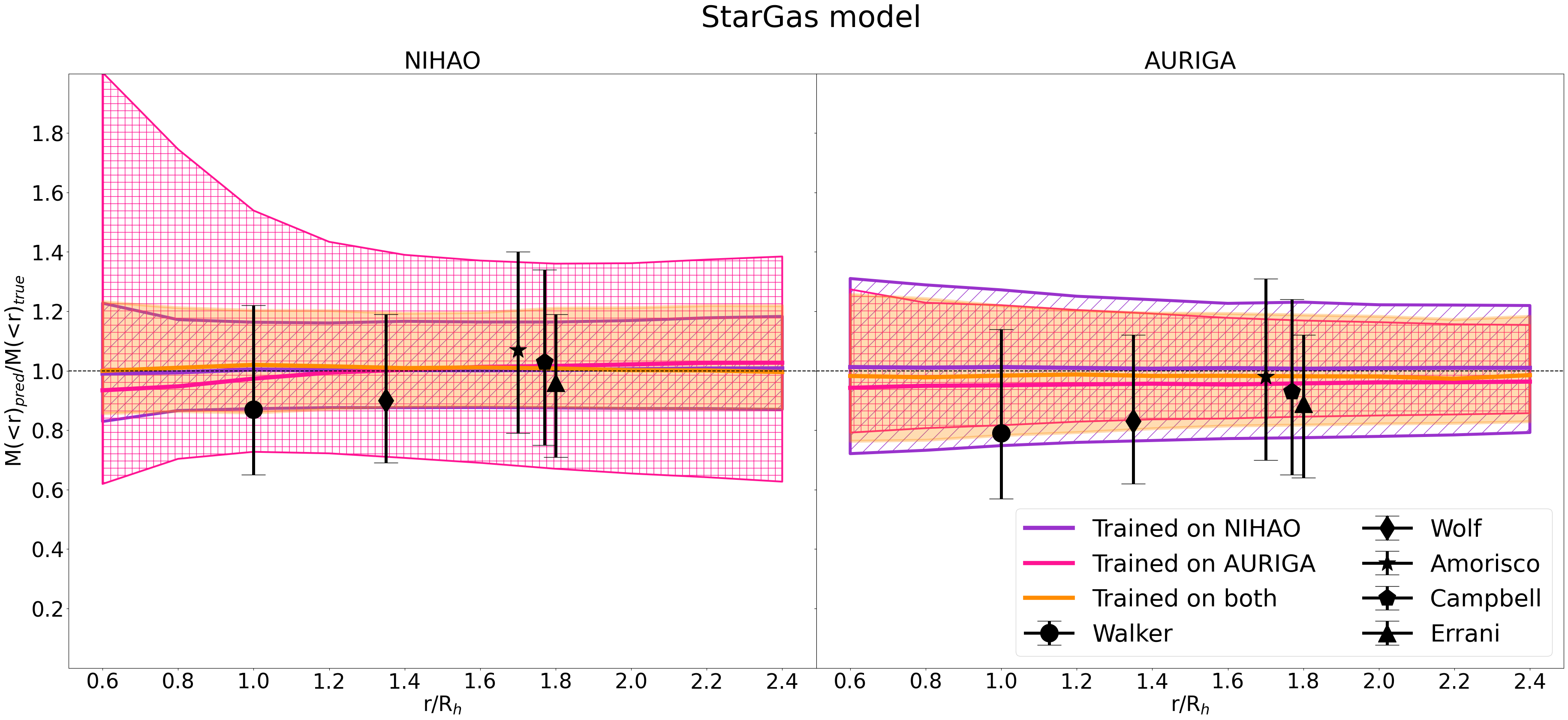}
    \caption{Ratio between the mass predicted by the neural network models and the real mass enclosed within different radii of the galaxies in the test sets of NIHAO and AURIGA galaxy projections. The colored lines show the median ratio using the mass estimated by the CNN for the training set, while the shadowed regions indicate the 1-$\sigma$. Purple: Results of the StarGas model trained on NIHAO galaxies. Pink: Results of the StarGas model trained on AURIGA galaxies. Orange: Results of the StarGas model trained on both simulation suites. The predicted to true enclosed mass ratios resulting from applying literature mass estimators to both NIHAO (left) and AURIGA (right) galaxies are shown as black symbols with 1-$\sigma$ errorbars.}
    \label{fig:CrossTesting}
\end{figure*}

Training and testing a neural network on a single suite of hydrodynamical simulations may lead to overfitting to the specific physical and numerical characteristics of that simulation set. While the model may perform well within that domain, its reliability on observational data, or even other simulations, can be compromised. Ideally, the network should learn general features in the input maps that are predictive of the mass profile, rather than adapting to the peculiarities of a particular simulation. A robust way to assess this generalization capability is through cross-simulation testing: training the model on one simulation suite and evaluating its performance on a different one, then comparing the results to those obtained when testing within the same suite.

We conduct a cross-testing analysis between the NIHAO and AURIGA simulation suites. A similar comparison was previously performed in \citet{sarrato25} using a model equivalent to our Star model, which revealed a clear limitation in generalization across simulations. In this work, we extend that analysis by examining how the inclusion of gas information influences the model’s ability to generalize between different simulation suites.

Fig. \ref{fig:CrossTesting} and Table \ref{tab:CrossTesting} presents the results of the cross-simulation analysis for the StarGas model. When the model is trained on both NIHAO and AURIGA simultaneously, its performance on each individual suite is comparable to that of models trained exclusively on the respective suite. In both cases, we observe a slight reduction in performance, which likely reflects a reduced emphasis on simulation-specific features and an improvement in the model’s generalization ability.

In contrast, when testing NIHAO galaxies with a model trained solely on AURIGA, the performance degrades severely, particularly at small radii. This failure may stem from the fact that approximately 90$\%$ of the NIHAO sample consists of dwarf satellite galaxies, whose inner mass profiles are strongly influenced by NIHAO’s feedback prescription, known to differ significantly from that used in AURIGA. As a result, the AURIGA trained model is likely unable to reproduce structural features such as the cores found in many NIHAO dwarfs, leading to large dispersion and systematic overestimation of the enclosed mass in the inner regions.

On the other hand, the model trained only on NIHAO performs consistently when applied to AURIGA galaxies. While the dispersion increases slightly relative to the AURIGA-trained model, the results remain stable, with only a mild bias toward underestimating the enclosed mass.

\section{Conclusions}
\label{sec:conc}

In this work, we developed and tested a probabilistic deep learning framework to infer the enclosed dynamical mass profiles of galaxies by combining stellar and gas kinematic information from realistic cosmological hydrodynamical simulations. Following previous work done by \citet{exposito23} and \citet{sarrato25}, our model is built around a multiple channel convolutional neural network (CNN) with a normalizing flow for uncertainty quantification, that uses projected stellar and gas 2D maps to predict the mass distribution across different radii.

By training on a diverse and physically motivated dataset drawn from the NIHAO simulation suites, we systematically explored the strengths and limitations of the approach. We used both mock HI observations using the MARTINI code \citep{oman19} and kinematic information from the star particles of the simulations. We compared variants of the model that use only stellar data, only gas data, or a combination of both, and we studied the effects of galaxy inclination on performance, as well as the biases of the model when applied to specific regions of the galaxies parameter space. We also trained and tested the model on the AURIGA simulation suites and we performed a cross-simulation testing of the model to study its generalization capabilities.

The main results of our study are:

\begin{itemize}
    \item The StarGas model, which integrates both stellar and gas input maps, consistently outperforms single-tracer models and standard literature mass estimators \citep{Walker_estim, Wolf_estim, Amorisco_estim, Campbell_estim, Errani_estim} when trained and tested on NIHAO galaxies. It reduces the scatter in mass profile predictions to a standard deviation of $\sim30\%$ across a range of radii and yields unbiased residuals at all scales (Fig. \ref{fig:MvsR}).

    \item The Star model suffers from strong biases related to galaxy inclination, significantly overestimating the mass in edge-on systems and underestimating it at intermediate inclinations. These projection effects are not effectively disentangled by the stellar channels alone. The Gas model avoids systematic inclination biases but shows greater dispersion, especially for edge-on systems. The StarGas model partially mitigates the inclination-related biases of the Star model and the increased uncertainty of the Gas model, demonstrating that combining both tracers provides complementary constraints on the gravitational potential (Fig. \ref{fig:Inc_and_RotDisp}).
    
    \item Cross-testing reveals that models trained on a single simulation suite are vulnerable to generalization issues due to differing feedback implementations and numerical effects. The StarGas model trained on AURIGA performs poorly when tested on NIHAO galaxies, with the gas implementation not outperforming previous results with a star-only simpler model \citep{sarrato25}, particularly in the inner regions. The bad performance is likely due to its inability to model cored mass profiles in dwarf galaxies, since AURIGA galaxies don't form cores. On the other hand, the model trained on NIHAO performs significantly better when tested on AURIGA galaxies. Training on a mixed dataset from multiple suites improves the model's generalization capacity, with only moderate increases in scatter. The model trained on both NIHAO and AURIGA achieves stable performance across simulation types. These results indicate that the StarGas model has better robustness to simulation-specific systematics (Fig. \ref{fig:CrossTesting}) than the star-only model from \citet{sarrato25}.
\end{itemize}

Overall, our results demonstrate the power of combining multiple kinematic tracers, such as HI and stellar maps, in a unified deep learning framework to infer galaxy mass profiles in a data-driven and probabilistic manner. Future work will focus on extending this model to observational datasets and further exploring the model’s interpretability and sensitivity to physical galaxy properties.

\begin{acknowledgements}
CB is supported by the Spanish Ministry of Science and Innovation (MICIU/FEDER) through research grant PID2021-122603NBC22. ADC is supported by the Agencia Estatal de Investigación, under the 2023 call for Ayudas para Incentivar la Consolidación Investigadora, grant number CNS2023-144669, proyecto "TINY". The authors wish to acknowledge the contribution of the IAC High-Performance Computing support team and hardware facilities to the results of this research. The freely available softwares \textit{pynbody} \citep{Pontzen13} and \textit{MARTINI} \citep{oman19} have been used for part of this analysis. The authors thank the AURIGA projects' PIs for making their data publicly available at https://wwwmpa.mpa-garching.mpg.de/auriga/data.html.
\end{acknowledgements}

\bibliographystyle{aa}
\bibliography{bib.bib}

\end{document}